\documentclass[epj]{svjour}
\usepackage{graphics}
\usepackage{epsfig}
\usepackage{textcomp}
\usepackage{amssymb}
\usepackage{amsmath}
\usepackage{mathptmx}
\DeclareSymbolFont{newfont}{OML}{cmm}{m}{it}
\DeclareMathSymbol{\Varrho}{3}{newfont}{37}
\begin{document}
\title{Modeling Alignment Enhancement for Solid Polarized Targets}
\author{D. Keller\inst{1}
}                     
\newcommand{\bra}[1]{\left\langle #1 \right|}
\newcommand{\ket}[1]{\left| #1 \right\rangle}

\offprints{}          
\institute{University of Virginia, Charlottesville, Virginia 22901}
\date{\today}
%
\abstract{
A model of dynamic orientation using optimized radiofrequency (RF) irradiation produced perpendicular to the holding field is developed 
for the spin-1 system required for tensor-polarized fixed-target experiments. The derivation applies to RF produced close to
the Larmor frequency of the nucleus and requires the electron spin-resonance linewidth to be much smaller than the nuclear
magnetic resonance frequency.
The rate equations are solved numerically to study a semi-saturated steady-state resulting from the two sources of irradiation: microwave from the DNP
process and the additional RF used to manipulate the tensor polarization.
The steady-state condition and continuous-wave NMR lineshape are found that optimize the spin-1 alignment in the polycrystalline materials used
as solid polarized targets in charged-beam nuclear and particle physics experiments.
}

\maketitle

\section{Introduction}
\label{intro}

Recent growing interest in tensor-polarized observables accessible in fix-target experiments stems from their unique ability to allow access to information not available
using a vector-polarized target.  The tensor-polarized structure functions \cite{jaffe} arising in spin-1 decomposition of the hadronic tensor in deep inelastic scattering (DIS) is one example of this.  An experiment to measure the leading-twist spin-1 tensor-polarized structure functions from a solid polarized target has been proposed at Jefferson Lab \cite{JP}, and it is also possible to study the effects of deuteron alignment in short range correlations \cite{Day,Duke}.  However, very little data on tensor-polarized observables from solid polarized target experiments exists to date \cite{smith1,smith2,otter1,smith3,otter2,boden} largely because it is necessary to maximize both the tensor polarization and beam intensity simultaneously to achieve sufficiently high measurement precision.

Maximizing the likelihood of interaction when probing tensor-polarized observables requires generating a high tensor polarization of the target.
The figure-of-merit for such experiments typically scales as the square of the tensor polarization.
In addition, time-dependent drifts are suppressed by the magnitude of the polarization, setting the scale of the asymmetry that the experiment can probe.  The requirement of having both high tensor polarization and beam intensity simultaneously has not been available in previous experiments.  Deuteron vector polarizations of 80-90\%, corresponding to tensor polarizations of 55-75\%, have been demonstrated in deuterated butanol and propanediol samples dynamically polarized at temperatures below 300 mK using $^3$He/$^4$He dilution refrigerators \cite{bofrost,frost}.  However, beam heating and radiation damage limit these systems to luminosities less than $10^{32}$ cm$^{-2}$ s$^{-1}$.  Much higher luminosities, up to $10^{36}$ cm$^{-2}$ s$^{-1}$, can be achieved with samples of ND$_3$ or $^6$LiD polarized at 1~K \cite{crabb1}, but in these cases the deuteron vector polarization is typically less than 50\%, corresponding to a tensor polarization less than 20\%.

This article examines a method whereby the tensor polarization can be significantly increased in 1 K systems by using RF radiation (hole burning) to manipulate the NMR line of deuterons in ND$_3$.  Tensor polarization manipulation using saturating RF has been discussed by previous authors \cite{schil,del} for ND$_3$ and 1,2-propanediol respectively.  The work in \cite{schil} suggested that such a method could be use to enhance tensor polarized targets in particle physics experiments but could only offer rough estimates of the tensor polarization after applying RF.  The work in \cite{del} offered a simple measurement technique using 1,2-propanediol samples dynamically polarized at 2.5 T and 0.3 K.  The RF application occurred only after switching off the microwave source for dynamic polarization and placing
the sample in the frozen spin condition at either 1.25 or 2.5 T and temperatures below 100 mK.  The authors found their determination of the polarization after RF manipulation inconsistent with experimental results found by measuring the tensor analyzing power $T_{20}$ in $\pi d$ elastic scattering.  The present work utilizes a model resulting in a measurement method based on numerical solutions to the spin-1 solid-effect rate equations that describe the three spin states of the deuteron under the dual processes of dynamic polarization and RF manipulation.  These solutions are then applied to data on ND$_3$ polarized at 1 K and 5 T, demonstrating an RF manipulation and measurement technique testable with scattering experiments.

The remainder of this article is organized as follows. In Sections 2 and 3 brief
descriptions are presented, respectively, of the deuteron lineshape of ND$_3$ and
enhancement with the intent to increase the deuteron tensor polarization. Section
4 gives a set of coupled differential equations that describe the action of the lineshape
during both dynamic polarization and RF saturation along with the bulk
behavior of the material. Details of the CW-NMR setup used to obtain the data needed
to parameterize the simulations are given in Section 5. Section 6 contains the
solutions and interpretation of the aforementioned rate equations. Further discussion
is given in Section 7 on the optimization of the tensor-enhancement mechanism and
resulting theoretical lineshape based on the simulation. Section 8 discusses
possible methods to verify the results using specialized NMR experiments as well as
the use of known tensor analyzing powers in nuclear scattering. 

\section{The Deuteron NMR Lineshape}
\label{signal}
The polarized target apparatus includes a strong magnet producing a cylindrically symmetric field $B_0$, oriented in a direction
taken as the $z$-axis.  In this case, the density matrix describing 
an ensemble of spin-1 particles like the deuteron, respects the same 
cylindrical symmetry and can be expressed as
\begin{eqnarray}
\Varrho&=&\frac{1}{3}\bold{1}+\frac{1}{2}P_nI_z+\frac{1}{6}Q_nI_{zz}.
\label{one}
\end{eqnarray}
Here, $I_z$ and $I_{zz}$ are the two remaining, nonzero Hermitian basis operators 
and pertain to rotations about the $z$-axis only. The parameters $P_n$ and $Q_n$
are the vector and tensor polarizations of the ensemble:
\begin{eqnarray}
P_n&=&<I_z> =\frac{n_+-n_-}{n}\mathrm{\hspace{1em} and}\\
Q_n&=&<I_{zz}>=\frac{1-3n_0}{n}\mathrm{,}\\\nonumber
\label{two}
\end{eqnarray}
where $n_i$ is the population of the $m_i$ magnetic substate and the total 
number of spins is $n=n_+ + n_0 + n_-$.  The polarizations may take values $-1 
\le P_n \le 1$ and $-2 \le Q \le 1$.  Note that the tensor polarization
(sometimes referred to as the alignment or quadrupole polarization) can 
be seen as a measure of the $n_0$ population: to increase the tensor 
polarization means decreasing $n_0$, and vice versa.  Because dynamic 
polarization requires doping the sample with free or unpaired electron
spins in the form of paramagnetic radicals, the polarization of these spins is also
specified:
\begin{eqnarray}
P_e&=&\frac{(n^e_+-n^e_-)}{n^e}.
\label{three}
\end{eqnarray}

When the Zeeman levels are populated according to Boltzmann statistics at some temperature $T$, the polarizations for a spin-1 system can be calculated using standard Brillouin functions:
\begin{eqnarray}
\label{four}
P_n&=&\frac{4\tanh (\hbar \omega_D/2kT)}{3+\tanh^2(\hbar \omega_D B_0/2kT)}\mathrm{,}\\\nonumber
Q_n&=&\frac{4\tanh^2 (\hbar \omega_D/2kT)}{3+\tanh^2(\hbar \omega_D B_0/2kT)}\mathrm{,\hspace{1em} and}\\\nonumber
P_e&=&\tanh(\hbar\omega^S_0B_0/2kT)\mathrm{,}
\end{eqnarray}
where $\hbar\omega_D(\hbar\omega^S_0)$ is the deuteron (electron) Zeeman energy.  
For the case where the system is in Boltzmann equilibrium, $Q_n$ only exists in the range 
from $0\leq Q_n \leq 1$, and the following relationship holds for $P_n$ and $Q_n$,
\begin{eqnarray}
Q_n&=&2-\sqrt{4-3P_n^2}.
\label{relation}
\end{eqnarray}
With only irradiation from the microwave source the DNP process can be described as the cooling of the nuclear moments to a very low spin temperature that can be 
either positive or negative, depending on the sign of the vector polarization.

The equilibrium electronic polarization $P_0$ comes directly from spin-temperature theory \cite{borghini},
$$P_0=-\textrm{tanh}(h\omega^S_0/2kT)\equiv-(1-r)/(1+r),$$
\begin{equation}
r\equiv exp(-h\omega^S_0/kT),
\end{equation}
where $T$ is the temperature and $k$ is the Boltzmann constant.  The value of $P_0$ is sensitive to the spin-temperature conditions imposed by the cryostat and RF environment and is treated as a parameter in the rate equations.  The deuteron nuclear equilibrium polarization is comparatively small for relevant field and temperature.

In this situation, the relation in Eq. \ref{relation} between 
$Q_n$ and $P_n$ still holds, and a measurement of the vector polarization can be used to determine the tensor polarization.
\begin{figure}
\begin{center}
\includegraphics[height=45mm, angle=0]{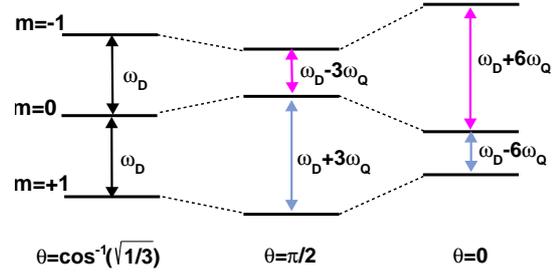}
\end{center}
\caption{The energy level diagram (not to scale) for deuterons in a magnetic field for three values of $\theta$ where $\hbar\omega_D$ is the deuteron Zeeman energy, $\hbar\omega_Q=(eq)(eQ)/8$ is the quadrupole energy, and $\theta$ is the angle between the magnetic field and the EFG.  The red (blue) lines indicate the transitions from the magnetic sublevels $-1\leftrightarrow0$ ($0\leftrightarrow1$).}
\label{levels}
\end{figure}
In addition to their magnetic dipole moment, spin-1 nuclei can have an electric quadrupole moment ($eQ$) which results from a nonspherical charge
distribution in the nucleus.  In the case of materials with cubic symmetry, the electric field gradient ($eq$) is zero and the quadrupole nuclei (e.g. $^6$LiD and HD) behave similarly to the spin-1/2 nuclei. The two Zeeman transitions for the deuteron in an isotropic condition are degenerate, resulting in a single NMR peak
spectrum in the frequency domain. For materials without cubic symmetry (e.g. ND$_3$), the interaction of the $eQ$ with the electric field gradient (EFG) breaks the degeneracy of the
energy transitions, leading to two overlapping absorption lines in the NMR spectra, which indicates a quadrupolar splitting ($\Delta\nu_Q$). The magnitude
of the quadrupole splitting is dependent on the EFG, the size of the quadrupolar moment, and the relative orientation of the EFG with respect to 
the magnetic field $B_0$.  The energy levels, Fig. \ref{levels}, of this type of spin-1 system can be expressed as,
\begin{eqnarray}
E_m&=&-m\hbar\omega_D+\hbar\omega_Q[(3\cos^2\theta-1)\\\nonumber
&+&\eta\sin^2\theta\cos2\phi](3m^2-2),
\label{ee}
\end{eqnarray}
where $\hbar\omega_D$ is the deuteron Zeeman energy, $\hbar\omega_Q=(eq)(eQ)/8$ is the quadrupole energy, and $\theta$ is the angle between the magnetic field and the EFG. The azimuthal angle $\phi$ and asymmetry parameter $\eta$ are required for describing energies where the EFG is not symmetric about the bond axis \cite{abr1}. The two allowed transitions from the magnetic sublevels $(-1\leftrightarrow0)$ and $(0\leftrightarrow1)$ correspond to energy differences $(E_{-1}-E_0)$ and $(E_{0}-E_1)$, respectively.  There are two resonant frequencies for any given $\theta$ corresponding
to the $(0\leftrightarrow1)$ transition with intensity $I_+$ and the $(-1\leftrightarrow0)$ transition with intensity $I_-$.
A fitting technique of the spin-1 NMR signal which relies on the system maintaining the thermal equilibrium relationship between vector and tensor polarization has been previously outlined \cite{dulya}. 
A spin-1 NMR signal is shown in Fig. \ref{fit} with a fit demonstrating the two intensities $I_+$ (in blue) and $I_-$ (in red).

It is convenient to define a dimensionless position in the NMR line
$R=(\omega-\omega_D)/3\omega_Q$, spanning the domain of the NMR signal.
The degree of axial symmetry and dependence on the polar angle can be understood by using the basis lineshape for an isotropic rigid solid or the Pake doublet \cite{pake}, allowing for a fit of the NMR data.  This is simply a sum of the areas of the two absorption lines.
The peaks of the Pake doublet ($R\sim\pm1$) correspond to the principal axis of the coupling interaction perpendicular ($\theta=\pi/2$) to the magnetic field. This is the most probable configuration within each transition, as indicated by the intensity of each peak. The opposing end in each transition in the lineshape corresponds to the configuration occuring when the principal axis of the coupling interaction is parallel ($\theta=0$) to the magnetic field; this has much less statistical significance, as indicated by the intensities in each transition around ($R\sim\mp2$).

The spin-spin interactions cause a distribution in the local holding field for a given spin, leading to slight variations of $\omega_D$. The spectrum displays 
a dipolar broadening from these effects, approximated by using a convolution of the density of states with a Lorentzian \cite{abr1,dulya}.
The resulting analytic function is defined over all values of $R$ and expressed as
{\small{
\begin{eqnarray}
\label{function}
\mathcal{F}&=&\frac{1}{2\pi \mathcal{X}}\left[2\cos(\alpha/2)\left(\arctan\left(\frac{\mathcal{Y}^2-\mathcal{X}^2}{2\mathcal{Y}\mathcal{X}\sin(\alpha/2)}\right)+\frac{\pi}{2}\right)\right.\\\nonumber
&+&\left.\sin(\alpha/2)\ln\left(\frac{\mathcal{Y}^2+\mathcal{X}^2+2\mathcal{Y}\mathcal{X}\cos(\alpha/2)}{\mathcal{Y}^2+\mathcal{X}^2-2\mathcal{Y}\mathcal{X}\cos(\alpha/2)}\right)\right],
\end{eqnarray}
}
where $\mathcal{X}^2=\sqrt{\Gamma^2+(1-\epsilon R-\eta\cos 2\phi)^2}$, $\mathcal{Y}=\sqrt{3-\eta\cos 2\phi}$ and $\cos\alpha=(1-\epsilon R-\eta\cos 2\phi)/\mathcal{X}^2$.  After a $\phi$-average and fit to ND$_3$ experimental data, $\eta\cos 2\phi$ evaluates to $\sim$0.04.  The Lorentzian width, $\Gamma$$\sim$0.05, is related to the degree of dipolar broadening of the NMR signal \cite{dulya} and $\epsilon$ references the specific intensity curve so that $\mathcal{F}(\epsilon=\pm1)=I_{\pm}$.

The NMR signal, properly fitted,
measures the intensity of each peak in the doublet.  These fitted intensities are used to extract the averaged polarization of the ensemble over the course of the nuclear scattering experiment.
\begin{figure}
\begin{center}
\includegraphics[height=65mm, angle=0]{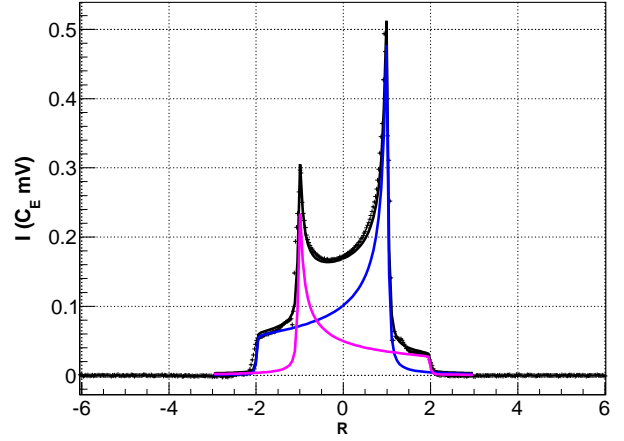}
\end{center}
\caption{A demonstration fit to ND$_3$ NMR data using the discussed lineshape. The fit based on the intensity from each peak is $P_n=42.3\%$ (vector polarization) and $Q_n=13.1\%$ (tensor polarization).  The intensity $I_+$ ($I_-$) is shown in blue (red). }
\label{fit}
\end{figure}

The response of a spin system to NMR probing RF irradiation is described by its magnetic susceptibility $\chi(R)=\chi'(R)-\chi''(R)$, where $\chi''$ is the absorption function and $\chi'$ is the dispersion function. The vector polarization of the spin-1 system with residual quadrupole coupling can be described \cite{goldman} by
\begin{eqnarray}\nonumber
P_n&=&\frac{2\hbar }{g^2\mu^2_N\pi N}\int_{-\infty}^{\infty}\frac{3\omega_Q\omega_D}{3R\omega_Q+\omega_D}\chi''(R)dR\\
&=&\frac{1}{C_E}\int_{-\infty}^{\infty} I_+(R)+I_-(R)dR,
\end{eqnarray}
where $g$ is the dimensionless magnetic moment of the particle with spin and $\mu_N$ is the nuclear magneton.
The total deuteron NMR signal, consisting of the output voltage integrated over frequency, is the total area of the two separate intensities $I_+$ and $I_-$ originating from the $(1\to0)$ and $(0\to-1)$ transitions, respectively. Each transition's intensity is proportional to the differences of the level populations within that transition so that $I_+=C_E(\rho_+-\rho_0)$ and $I_-=C_E(\rho_0-\rho_-)$, where $C_E$ is the calibration constant encoding the relationship between signal area and polarization and $\rho_i$ is the level population normalized to the total ensemble population. A measurement with the spin system in a known polarization state determines the calibration constant, such as when the ensemble is in thermal equilibrium with the lattice and Eq. \ref{four} can be used.  After acquiring the calibration, the vector polarization can be calculated as $P_n=(I_++I_-)/C_E$, and the tensor polarization as $Q_n=(I_+-I_-)/C_E$ for RF manipulated or dynamically enhanced targets.

In the example fit to the ND$_3$ NMR data (Fig. \ref{fit}), the magnitude of the powder pattern, $I$, is in units of millivolts multiplied by the calibration constant $C_E$. Based on the intensity from each peak, a measurement of $P_n=42.3\%$ (vector polarization) and $Q_n=13.1\%$ (tensor polarization) results.

\begin{figure}
\begin{center}
\includegraphics[height=55mm, angle=0]{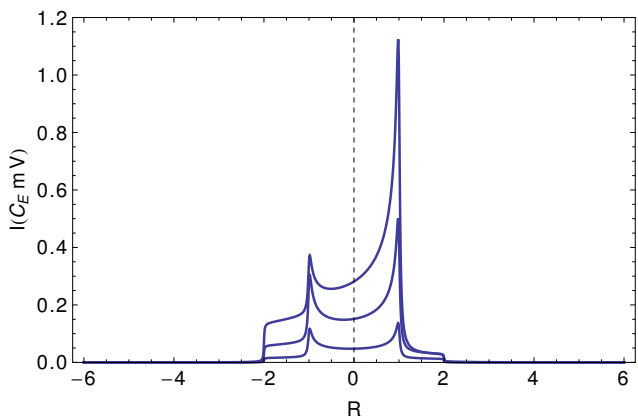}
\end{center}
\caption{Simulated lineshape showing the increase in positive vector polarization of 13\%, 42\%, and 78\%.}
\label{lines}
\end{figure}

As the sample vector polarization increases, the difference between the intensities $I_+$ and $I_-$ becomes greater.
The change to the theoretical lineshape for increasing positive vector polarization is shown in Fig. \ref{lines} for polarization of 13\%, 42\%, and 78\%. The area of the NMR signal gets progressively larger with 
increasing polarization. Understanding the change in the relative size of the intensities $I_+$ and $I_-$ as a function of vector polarization is critical for the RF tensor enhancement. As the vector polarization decreases, the degree of potential tensor polarization enhancement per total area of NMR signal increases. During scattering experiments the relative improvement in tensor enhancement at low polarization can partially offset the loss of polarization due to radiation damage. 


\section{Polarization Enhancement}

\label{approach}

To manipulate the magnitude of tensor polarization during DNP pumping, a separate source of coil-generated RF irradiation is used to selectively saturate some
portion of the deuteron NMR line. Application of RF irradiation at a single frequency or over a frequency range induces transitions between the
magnetic sublevels within the frequency domain of the applied RF. A spin-diffusion rate that is
small compared to the effective nuclear relaxation rate allows for significant changes to the NMR line via the RF, which can be strategically used
to manipulate the spin-1 alignment. The present study gives the rate equations for both the microwave and the additional coil-RF irradiation, solving them numerically for the polarizations of an electron spin-1 nuclear system.  A semi-saturated steady-state condition is proposed which
manipulates and holds the magnetic sublevels responsible for polarization enhancement. To be useful for a nuclear experiment setting, the target
ensemble averaged tensor polarization must be increased and held in a continuous mode.
This model of the spin dynamics produces a simple simulation which generates the CW-NMR lineshape that maximizes tensor polarization values for the optimal steady-state condition.

Manipulation of the population in the $m=0$ state through the use of RF irradiation at certain positions in the doublet results in limited modular control
of the target alignment. Selective excitation of one of the allowed transitions over a frequency range reduces the absolute value of tensor polarization at the RF frequencies but enhances it at the same polar angle $\theta$ in the other transition.
The RF sweep rate must be fast with respect to the relaxation rates so as to simultaneously manipulate a portion of the NMR line. The irradiation from the RF
causes population exchange between the two sublevels in a transition, transforming pure Zeeman order into a combination of Zeeman and quadrupolar order. The other transition responds reflecting the change in population at $\theta$ and leading to an enhancement through the polarization and relaxation pathways generated by the RF. 

The deuterated ammonia target material possesses one of the highest deuteron polarizations along with good resistance to radiation damage under a charged beam
\cite{uva,bonn}, making it the essential spin-1 solid target for electron/proton scattering (such as in DIS and the proton-deuteron Drell-Yan process). The $^6$LiD target material has considerable radiation resistance; however, its cubic symmetry results in an NMR spectrum with no first order quadrupole splitting, making it inadequate for tensor polarization enhanced scattering experiments. Under optimal conditions in a high-cooling power evaporation refrigerator, ND$_3$ can be polarized to well above 45\% at 5 T. The polarization mechanism for radiation doped $^{14}$ND$_3$ in various thermal conditions is complex and still under study. Measurements of the ESR line of {\it warm} ($\sim$87 K) irradiated ND$_3$ indicate that radical $\dot{\text{N}}$D$_2$ is responsible for the DNP process \cite{meyer2}.
This warm dose is typically in the form of 10-20~MeV electrons, with integrated fluxes of about $10^{17}$ e$^-$/cm$^2$. 
The Larmor frequency of $^{14}$N is small with respect to the ESR linewidth of the radiation-induced paramagnetic centers; however, for the Larmor frequency of the deuteron this is less true \cite{goertz,meyer2}. The theory of equal spin temperature does not apply to ND$_3$, and there is some evidence \cite{crabb1,meyer2,meyer3} that the differential solid-effect \cite{abr1,leifson,abr2} is the dominant spin-transfer mechanism for samples prepared in this manner.
However, ND$_3$ exhibits unusual behavior as it accumulates additional {\it cold} ($\sim$1 K) irradiation during the scattering experiment.
First, the maximum attainable polarization increases significantly, from about 15\% to nearly 50\%.  Second, the optimum microwave frequencies for positive and negative polarization separate \cite{meyer3,mckee}.  A full understanding of all aspects of the observed behavior of irradiated ND$_3$ and its polarization mechanisms requires much more research.  In this initial study, the solid-effect is used demonstrating a technique to model the RF-manipulated NMR line with sensitivity to the polarization and relaxation pathways.  The microwaves are assumed to be at resonance for the material under study.

The process of the solid-effect uses unpaired electrons in free atoms to interact with a nuclear spin. These spins are subject to a static magnetic field $\mathbf{B_0}$ at cryogenic temperature with the system receiving irradiation from a microwave field $2\textbf{B}_{\mu}\textrm{cos}\omega_{\mu} t$ at frequencies that induce simultaneous ESR and NMR. The electrons are quickly polarized by brute force in the strong holding field while the vector polarization comes from the DNP process. An additional field is imposed for the purpose of manipulating the spin-1 alignment through use of a modulated RF field $2\textbf{B}_{\nu}\textrm{cos}\omega_{\nu} t$ produced by current in a helical wire coil around the coupled spin system. The frequency $\omega_{\mu}$ is chosen to be near the ESR frequency $\omega^S_0=-\gamma_s|B_0|$ and $\omega_{\nu}$ is chosen to be at the Larmor frequency $\omega^I_0=\gamma_I|B_0|$, where $\gamma_S$ and $\gamma_I$ are the gyromagnetic ratios of the electron and nucleon spin respectively. Choosing the $z$-axis to be the direction of the holding field, as in Eq. \ref{one}, the single-frequency solid-effect Hamiltonian contains only the Zeeman, RF-irradiation, and hyperfine contributions to the system so that,
\begin{eqnarray}
\mathcal{H}_{SE}&=&\omega^S_{0}S_z-\omega^I_{0}I_z\\\nonumber
&+&2\omega^S_{\mu}S_x\textrm{cos} \omega_{\mu} t+2\omega^I_{\nu}I_y\textrm{cos} \omega_{\nu} t +\textbf{S}\cdot \textbf{A} \cdot \textbf{I},
\label{H}
\end{eqnarray}
where $\omega^S_{\mu}=\gamma_s|\textbf{B}_{\mu}|$ and $\omega^I_{\nu}=\gamma_I|\textbf{B}_{\nu}|$, such that $\gamma_s$ is negative while $\gamma_I$ is positive. The set of dot products represents the hyperfine interaction between the two spins where $\textbf{A}$ is the hyperfine interaction tensor.

The microwave term can induce simultaneous transitions of the electron spin and the nuclear spin, the RF-irradiation term can only affect nuclear transitions. The two Zeeman terms in the Hamiltonian can be interpreted as the unperturbed part, and the RF terms with the hyperfine interaction term can be interpreted as the perturbation. The resulting rate of the transitions induced by the microwave takes the form
\begin{equation}
\beta^{\pm}=2\pi\left(\frac{\textbf{B}_{\mu}a_{\mu}}{\textbf{B}_0 }\right)^2\delta(\omega^S_0\pm\omega^I_0-\omega_\mu),
\end{equation}
which reflects the two types of flip-flop transitions producing both positive and negative polarization. The resulting rate of the transitions induced by the RF takes the form
\begin{equation}
\xi=2\pi\left(\frac{\textbf{B}_{\nu}a_{\nu}}{\textbf{B}_0 }\right)^2\delta(\omega^I_0-\omega_{\nu}),
\label{six}
\end{equation}
which can only play a role between nuclear spin energy levels that differ by the spin of the mediating photon while the electron spin in the hyperfine coupling remains unchanged. In each expression, $a_{\mu}$ and $a_{\nu}$ are constants depending on the coupling strength.

The DNP polarization enhancement process is modeled with the solid-effect while the tensor polarization enhancement
process is modeled with dynamic orientation using the coil-RF at a single frequency in the CW-NMR signal.  A family of numerical solutions are generated and used to interpret the behavior of the magnetic sublevel population at the coil-RF irradiated frequency.

\section{Modeling the Lineshape}
\label{rates}
The method presented here generates a theoretical lineshape based on the rate equations at just a single line position, $\mathcal{R}$, where the RF-coil irradiation is selectively saturated or semi-saturated.
These solutions combine with bulk spin effects acquired empirically to describe the dynamics of the overlapping absorption lines in the CW-NMR.
The relationship between the two intensities $I_+$ and $I_-$ at $\mathcal{R}$ must be calculated in order to obtain the overall area change under each absorption line which can then be used to calculate the resulting vector and tensor polarization.

A simulation is used to describe any particular CW-NMR frequency response of the longitudinal magnetization behavior under the implemented RF-irradiation.  The physical environment of the spin system is not rigorously simulated here rather, a simplified approach is used which relies on empirical information extracted from NMR and polarization build up data.  The lineshape in the model pertains to frozen polycrystalline ND$_3$ with free radicals in a glassy matrix.  At the required temperature of $\sim$1 K, there are little relevant molecular dynamics with rotations mostly quenched by lockup from intermolecular bonding.
The simulation uses the lineshape from Eq. \ref{function}, augmented with Voigt functions, as a response to
the coil-RF irradiation.  Here, 500 bins represent the 500 RF scan steps of the CW-NMR.
Monte Carlo sampling of a distribution defined by the theoretical lineshape generates the values of the 500 bins.
Iteratively regenerating the distribution as the system evolves under the correct RF conditions simulates changes in the system.
Parameterization of the relationship between microwave power and the material characteristics requires empirical data; the simulated lineshape can then be used to study the target alignment phenomenology for the RF-manipulated CW-NMR.

Adding a measured diffusion parameter related to the NMR line recovery rate (after RF terminates) incorporates the transverse effect of homonuclear polarization-transfer (spin-diffusion).  A fit also parameterizes the spread of the RF power absorbed in the NMR-line. 
In the real system, polarization transfer through spin-diffusion occurs as part of the bulk material polarization process.  Spectral diffusion also plays a role in the bulk behavior of the material, as seen in the line recovery over time after a selective excitation.  Because the rate calculations here hold strictly for a single frequency, these aspects of spin-diffusion are largely ignored in the handling of the rate equations.

The lineshape changes as the intensity decreases at the RF line position $R=\mathcal{R}$ and increases for the same $\theta$ at $R=-\mathcal{R}$. Inhomogeneous broadening occurs
in individual spin packets within the absorption envelope of the line.  The application of RF at a select frequency $\mathcal{R}$ in the CW-NMR signal results in a range of resonance-frequency spins in the envelope experiencing different field strengths.  This leads to a change in the width of the individual Lorentzian spin packets in each absorption line in the irradiated
frequency domain. The variation in the Lorentzian widths depends on the inhomogeneity of the field generated by the RF. The lineshape depletion over the affected frequency range is approximated by a subtracted Voigt. The Lorentzian width at each spin packet depends on the spin-spin relaxation under the inhomogeneous field of the RF. The shape of the resulting Voigt represents the response of the absorption envelope to the power profile generated by the RF-coil.  These characteristics largely depend on the coil design and amplitude of the signal generated.  The region around $R=\mathcal{R}$ in each absorption line can be simulated by fitting a Voigt function to NMR data and using 
the resulting fit parameters in the simulation assuming equal Voigt widths for the two overlapping absorption lines.   

The lineshape also changes in the region around $R=-\mathcal{R}$, where the intensities increase by a positive Voigt function with initially the same shape as the negative Voigt at a smaller amplitude. An additional broadening occurs from the spin-diffusion to neighboring
spins.  The spin-diffusion process occurs when individual nuclear spins undergo continuous exchange of energy, resulting in a transfer of polarization between spins of close proximity.
The mechanism has a unitary quantum-mechanical nature with geometric dependence governing the fundamental rate of the flip-flop process and can be approximated using the diffusion equation \cite{abr1,bloembergen,goldman2,ernst}.
Changes occur to the line after selective excitation when nuclear spin polarization transfers across the NMR spectrum (spectral diffusion).
In the steady-state condition of dynamic orientation, using both DNP and the coil-generated RF, broadening from spin-diffusion competes with the continual RF pumping, leading to a measurable width that can parameterize the simulations.  The diffusion is more evident and easier to measure after discontinuing all sources of RF irradiation and allowing the enhancement at $R=-\mathcal{R}$ to evolve in time. Changes to the Gaussian component of the initial width of the positive Voigt continue in time such that $\sigma(t)=\sigma_0+\sigma_d\sqrt{t}$, where the polarization moves within the line at time $t$ with diffusion constant $\sigma_d$ \cite{bloembergen}. Neglecting diffusion, the Gaussian $\sigma$ of the positive Voigt is the same as it is for the negative Voigt $\sigma_0$. 
The overall change in the lineshape results from addition of the Voigt functions to the two absorption lines. Fig \ref{para} shows a fit used to extract the Voigt parameterization from ND$_3$ NMR data. The intensity change to each absorption line requires studying the rate equations at $R=\mathcal{R}$.  Understanding the
relationship between the intensities $I_+(R=\mathcal{R})$ and $I_-(R=\mathcal{R})$ gives a unique constraint to the central amplitude in each Voigt.  Given the shape of the Voigt (determined experimentally) only the rate equation solutions at this single position in the line $\mathcal{R}$ are needed to simulate the rest of the NMR signal and to generate a theoretical lineshape after any RF manipulation.

\begin{figure}
\begin{center}
\includegraphics[height=67mm, angle=0]{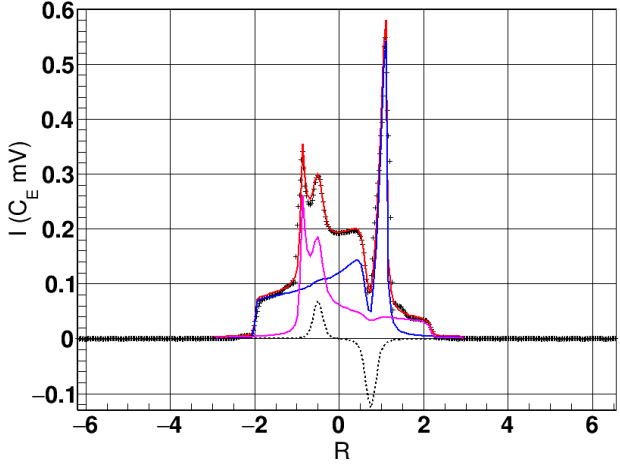}
\end{center}
\caption{Partial saturation and response of the $I_+$ transition in
polarized ND$_3$.  The RF is applied at a frequency corresponding to $R=0.75$,
which partially saturates the $I_+$ transition (blue curve) for deuteron
spins with $\theta$ near $73^{\circ}$.  The $I_-$ transition (red curve) is also
slightly affected, in this case for deuterons near $40^{\circ}$.  The
black line is the sum of both transitions, while the crosses represent
experimental NMR data points.  Note that saturation of one transition at $R=\mathcal{R}$
results in an increase in the other transition at $R=-\mathcal{R}$.  The dashed line
shows both the positive and negative contributions from each Voigt
function in the absorption line.}
\label{para}
\end{figure}

The solid-effect rate equations for spin-1 have been written and investigated in previous work \cite{jeffries,fedders,jizhi}, and a similar notation is used here.
All rates in the solid-effect equations \cite{jizhi} are preserved in the formalism that follows, but in the final evaluation only the irradiation-induced rates from DNP and the coil-generated RF prove critical to the dynamics of interest.
In the following derivation, an extension is made to include RF near the Larmor frequency to study the rates at a select position in the NMR line. 
The electronic and nuclear spin have magnetic quantum numbers $m_S=\pm1/2$ and $m_I=-1,0,1$, respectively. The rates are all expressed in terms of $\omega_1$,
the rate for the $\Delta m_S=\pm1$, $\Delta m_I=0$ transition.
The fundamental rate constant $\omega_1$ is defined as $\omega_1\equiv1/(2T_{1e})$, where $T_{1e}$ is the longitudinal relaxation rate for the electrons. The transition rate involving the microwave
pumping in the electron-spin system is $\omega_1\beta$ where $\Delta m_S=\pm1$ and $\Delta m_I=\pm1$.  The nuclear relaxation rates are $\omega_1\lambda$, where $\Delta m_S=0$ and $\Delta m_I=\pm1$ and $\omega_1\sigma$, where $\Delta m_S=\pm1$ and $\Delta m_I=\pm1$. The transition rate induced by the RF magnetic field is $\omega_1\xi$. Also included are two NMR relaxation rates incorporating minor polarization losses induced by the measurement process \cite{jizhi}. The rate $\phi_1 \omega_1$ is the single-spin transition due to $\Delta m_I=\pm1$ transitions, and the rate $\phi_2 \omega_1$ comes from the double quantum transition $\Delta m_I=\pm2$ transitions.  Neglecting the quadrupole interaction for the time being, the energy level diagram is shown in Fig. \ref{decay}.  All discussed rates are preserved in the derivation, but the RF transition rates $\omega_1\beta$ and $\omega_1\xi$ are dominant.  Many of the smaller rates are negligible in the final evaluation of the solutions.
\begin{figure}
\begin{center}
\includegraphics[height=65mm, angle=0]{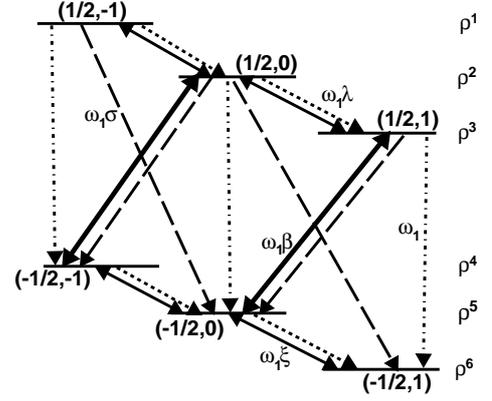}
\end{center}
\caption{Energy level diagram for the electronic and nuclear spins $(m_s,m_I)$ for the Zeeman states of an electron-deuteron system that is weakly coupled by the dipole-dipole interaction using DNP with an additional RF irradiation field. The rates are explained in the text.}
\label{decay}
\end{figure}

The population and orientation are a function of two distinct types of irradiation of field strengths $\textbf{B}_{\mu}$ and $\textbf{B}_{\nu}$. Considering the strength of
each field as an independent control parameter in the two decay constants, $\beta$ and $\xi$, respectively, leads to multiple states that are saturated, semi-saturated, or exhibit transient
behavior in which the dynamic population of the sublevels can only be understood by solving the rate equations for the system.

For a system of six levels with $S=\pm1/2$ and $I=1,0,-1$, let $\rho^i$ be the population and $E^i$ the energy of the $i$th state. Using the above-mentioned set of applied RF fields and relaxation processes, the change in the population for all energy levels is given by the rate equations,
\begin{equation}
2T_{1e}\frac{d\rho^i}{dt}=\sum_{j\neq i}\left(\rho^j\omega_{ji}-\rho^i\omega_{ij}\right),
\end{equation}
with $\omega_{ji}$ ($\omega_{ij}$) being the spin dependent coefficients for the gain (loss) pathways from state $j\to i$ ($i\to j$). Note that each coefficient of the relaxation terms represents
the decay constant for thermal processes and must contain the corresponding multiplicative factor for the pathways that follow a transition probability according to Boltzmann statistics. The irradiation-induced transition rate contributions to the magnetic sublevel populations are found by taking into account all of the irradiation-induced spin-flipping processes:
{\small{
\begin{eqnarray}\nonumber
\dot{\rho^1}(\textbf{B}_{\mu},\textbf{B}_{\nu}) & = & \xi \omega_1(\rho^2-\rho^1)\\\nonumber
\dot{\rho^2}(\textbf{B}_{\mu},\textbf{B}_{\nu}) & = & \beta \omega_1(\rho^4-\rho^2)+\xi \omega_1(\rho^3-2\rho^2+\rho^1)\\\nonumber
\dot{\rho^3}(\textbf{B}_{\mu},\textbf{B}_{\nu}) & = & \beta \omega_1(\rho^5-\rho^3)+\xi \omega_1(\rho^2-\rho^3)\\\nonumber
\dot{\rho^4}(\textbf{B}_{\mu},\textbf{B}_{\nu}) & = & \beta \omega_1(\rho^2-\rho^4)+\xi \omega_1(\rho^5-\rho^4)\\\nonumber
\dot{\rho^5}(\textbf{B}_{\mu},\textbf{B}_{\nu}) & = & \beta \omega_1(\rho^3-\rho^5)+\xi \omega_1(\rho^4-2\rho^5+\rho^6)\\\nonumber
\dot{\rho^6}(\textbf{B}_{\mu},\textbf{B}_{\nu}) & = & \xi \omega_1(\rho^5-\rho^6).\\\nonumber
\end{eqnarray}
}
The populations are normalized to unity so that $\sum_i\rho^i=1$. Expressing the full rate equations in terms of polarization yields,
{\small{
\begin{eqnarray}\nonumber
T'_{1e}\dot{P_e} & = &\frac{\beta}{6f_0}\left[P_eQ_n(\theta)-4P_e-3P_n(\theta)\right]\\\nonumber
& + & \frac{\sigma}{6f_0}(1+r)(P_e-P_0)Q_n(\theta)-(P_e-P_0)\\\nonumber
T'_{1n}\dot{P_n}(\theta) & = &-\frac{\beta}{4f_1}\left[P_n(\theta)+\frac{4}{3}P_e-\frac{1}{3}P_eQ_n(\theta)\right]\\\nonumber
& + &\frac{\lambda}{12f_1}(1+r)P_0(Q_n(\theta)-4)\\\nonumber
& - & \frac{\xi}{2f_1}P_n(\theta)-P_n(\theta) \\\nonumber
T'_{1q}\dot{Q_n}(\theta) & = &-\frac{3}{4f_2}\left[\beta\left(Q_n(\theta)+P_eP_n(\theta)\right)+\lambda(1+r)P_0P_n(\theta)\right]\\\nonumber
& - & \frac{3\xi}{2f_2}Q_n(\theta)-Q_n(\theta),\\
\label{rateqs}
\end{eqnarray}
}
where
\small{
\begin{eqnarray}\nonumber
f_0=\frac{(1+r)T_{1e}}{2T'_{1e}},~f_1=\frac{(1+r)T_{1e}}{2cT'_{1n}},~f_2=\frac{(1+r)T_{1e}}{2cT'_{1q}},
\end{eqnarray}
}
and the effective longitudinal relaxation rates are
\small{
\begin{eqnarray}\nonumber
T'_{1e} & = &T_{1e}\left(1+\frac{4}{3}\sigma\right)^{-1}\\\nonumber
T'_{1n} & = &T_{1e}\left(\phi_1+\frac{1}{2}c\lambda+\frac{1}{2}c\sigma(1-P_eP_0)\right)^{-1}\\\nonumber
T'_{1q} & = &T_{1e}\left(\phi_2+\frac{3}{2}c\lambda+\frac{1}{2}c\sigma(1-P_eP_0)\right)^{-1}\\\nonumber
\end{eqnarray}
}
The ratio of unpaired electrons to nuclear spins $c=n^e/n$ is a parameter used to related to the materials paramagnetic spin density. The polarizations are functions of $\theta$, the angle between the magnetic field and the electric field gradient. Expressing the rates in this way only allows us to understand the change in polarizations if the RF irradiation
was applied at the same $\theta$ in the line or over the entire resonance range. Applying RF to the same $\theta$ requires selective saturation at two frequencies in the NMR line simultaneously or at the center of the NMR line where $\theta$ is the same for both transitions ($\theta=54.7^{\circ}$).  Here our interest lies in applying RF at a single frequency to understand the changes in the overlapping absorption lines.

Modeling the RF-manipulated NMR line for the purpose of alignment enhancement requires rewriting the rate equations to study the change in polarizations at a single irradiated frequency. It is convenient to use the previously defined NMR line position, $R$, which spans the domain of the NMR signal. The rate contributions to the magnetic sublevel populations at the irradiated position, $R=\mathcal{R}$ for $\textbf{B}_{\nu}$ in the transition $(-1\to 0)$ are,
{\small
\begin{eqnarray}\nonumber
\dot{\rho^1}(\textbf{B}_{\nu},\mathcal{R}) & = & \xi \omega_1(\rho^2(\mathcal{R})-\rho^1(\mathcal{R}))\\\nonumber
\dot{\rho^2}(\textbf{B}_{\nu},\mathcal{R}) & = & \xi \omega_1(\rho^1(\mathcal{R})-\rho^2(\mathcal{R}))\\\nonumber
\dot{\rho^3}(\textbf{B}_{\nu},\mathcal{R}) & = & 0\\\nonumber
\dot{\rho^4}(\textbf{B}_{\nu},\mathcal{R}) & = & \xi \omega_1(\rho^5(\mathcal{R})-\rho^4(\mathcal{R}))\\\nonumber
\dot{\rho^5}(\textbf{B}_{\nu},\mathcal{R}) & = & \xi \omega_1(\rho^4(\mathcal{R})-\rho^5(\mathcal{R}))\\\nonumber
\dot{\rho^6}(\textbf{B}_{\nu},\mathcal{R}) & = & 0.\nonumber
\end{eqnarray}
}
These contributions are for positive polarization where the populations $\rho^2+\rho^5>\rho^1+\rho^4$. The same
can be expressed for the transition $(0\to 1)$,
{\small
\begin{eqnarray}\nonumber
\dot{\rho^1}(\textbf{B}_{\nu},\mathcal{R}) & = & 0\\\nonumber
\dot{\rho^2}(\textbf{B}_{\nu},\mathcal{R}) & = & \xi \omega_1(\rho^3(\mathcal{R})-\rho^2(\mathcal{R}))\\\nonumber
\dot{\rho^3}(\textbf{B}_{\nu},\mathcal{R}) & = & \xi \omega_1(\rho^2(\mathcal{R})-\rho^3(\mathcal{R}))\\\nonumber
\dot{\rho^4}(\textbf{B}_{\nu},\mathcal{R}) & = & 0\\\nonumber
\dot{\rho^5}(\textbf{B}_{\nu},\mathcal{R}) & = & \xi \omega_1(\rho^6(\mathcal{R})-\rho^5(\mathcal{R}))\\\nonumber
\dot{\rho^6}(\textbf{B}_{\nu},\mathcal{R}) & = & \xi \omega_1(\rho^5(\mathcal{R})-\rho^6(\mathcal{R})),\nonumber
\end{eqnarray}}
where these contributions are again used for positive polarization.
Here both sets of equations correspond to a different $\theta$ at the position $\mathcal{R}$ in the line. The normalization
$\sum_i\rho^i(\mathcal{R})=dn$ integrates over frequency to unity and represents the fractional magnetic sublevel occupancy at $\mathcal{R}$.  The rate equations can then be expressed using the vector polarization $P_n$ and tensor polarization $Q_n$, which are unique to
$\mathcal{R}$ but are coupled to the vector polarization $\mathcal{P}_n$ and tensor polarization $\mathcal{Q}_n$ at the same polar angle $\theta$ with negative $\mathcal{R}$.
The final rate equations at the RF irradiated position $\mathcal{R}$ in the NMR line are,
{\small
\begin{eqnarray}\nonumber
T'_{1e}\dot{P_e} & = &\frac{\beta}{6f_0}\left[g_1P_eQ_n-4P_e-3g_2P_n\right]\\\nonumber
& + & \frac{g_1\sigma}{6f_0}(1+r)(P_e-P_0)Q_n-(P_e-P_0)\\\nonumber
T'_{1n}\dot{P_n} & = &-\frac{\beta}{4g_2f_1}\left[g_2P_n+\frac{4}{3}P_e-\frac{g_1}{3}P_eQ_n\right]\\\nonumber
& + &\frac{g_1\lambda}{12g_2f_1}(1+r)P_0(Q_n-\frac{1}{g_1}4)\\\nonumber
& - & \frac{\xi}{3f_1}\left[\frac{3}{2}(\mathcal{P}_n+1)-\frac{1}{2}(1-\mathcal{Q}_n)+(1-Q_n)\right] \\\nonumber
& - &P_n \\\nonumber
T'_{1q}\dot{Q_n} & = &-\frac{3g_2}{4g_1f_2}\left[\beta\left(\frac{g_1}{g_2}Q_n+P_eP_n\right)+\lambda(1+r)P_0P_n\right]\\\nonumber
& - & \frac{\xi}{3f_2}\left[\frac{3}{2}(\mathcal{P}_n+1)-\frac{1}{2}(1-\mathcal{Q}_n)-(1-Q_n)  \right] \\\nonumber
& - & Q_n\\\nonumber
T'_{1e}\dot{P_e} & = &\frac{\beta}{6f_0}\left[g_3P_e\mathcal{Q}_n-4P_e-3g_4\mathcal{P}_n\right]\\\nonumber
& + & \frac{g_3\sigma}{6f_0}(1+r)(P_e-P_0)\mathcal{Q}_n-(P_e-P_0)\\\nonumber
T'_{1n}\dot{\mathcal{P}_n} & = &-\frac{\beta}{4g_4f_1}\left[g_4\mathcal{P}_n+\frac{4}{3}P_e-\frac{g_3}{3}P_e\mathcal{Q}_n\right]\\\nonumber
& + &\frac{g_3\lambda}{12g_4f_1}(1+r)P_0(\mathcal{Q}_n-\frac{1}{g_3}4)\\\nonumber
& + & \frac{\xi}{6f_1}\left[\frac{3}{2}(\mathcal{P}_n+1)-\frac{1}{2}(1-\mathcal{Q}_n)+(1-Q_n)\right] \\\nonumber
& - &\mathcal{P}_n \\\nonumber
T'_{1q}\dot{\mathcal{Q}_n} & = &-\frac{3g_4}{4g_3f_2}\left[\beta\left(\frac{g_3}{g_4}\mathcal{Q}_n+P_e\mathcal{P}_n\right)+\lambda(1+r)P_0\mathcal{P}_n\right]\\\nonumber
& - & \frac{\xi}{6f_2}\left[\frac{3}{2}(\mathcal{P}_n+1)-\frac{1}{2}(1-\mathcal{Q}_n)-(1-Q_n)  \right] \\\nonumber
& - & \mathcal{Q}_n.\\
\label{rateqs2}
\end{eqnarray}
}
The $g_i$ are constants defining the normalization at position $\mathcal{R}$ in the line. The normalization constants for $Q_n$ and $P_n$ are $g_1$ and $g_2$, respectively.
Likewise, the normalization constants for $\mathcal{Q}_n$ and $\mathcal{P}_n$ are $g_3$ and $g_4$, respectively.  The normalization constants characterize the initial polarizations for a given NMR line position. The set of equations shown in Eq. \ref{rateqs2} approximate the behavior of both overlapping absorption lines under the irradiation field $\textbf{B}_{\nu}$ with the microwaves for DNP active.

\section{Empirical Techniques}
Experimental CW-NMR data is used to parameterize the simulation as well as demonstrate qualitatively that the predictions overlap well with NMR experiments.
All data used to compare with the calculations presented in this work were taken at the University of Virginia's Solid Polarized Target Lab. The experimental data is taken
at 5 T and 1 K with an evaporation refrigerator \cite{crabb1,crabb2} that has a cooling power of just over 1 W and a DNP microwave source generating approximately 20 W. This microwave
power is reduced to less than 1 W after attenuation to reach the target cell.  The cell is 2.5 cm in diameter and 3 cm long which holds about 10 g of ND$_3$.  The nuclear spin polarization is measured with an NMR coil and Liverpool Q-meter \cite{nmr}. The RF
susceptibility of the material is inductively coupled to the NMR coil, part of a series LCR Q-meter circuit tuned to the Larmor frequency of the nuclei of interest.
This provides a non-destructive CW-NMR.  For selective excitation, an additional coil for RF-irradiation is connected to a generator and amplified to around 10 mW.  There
are many variations of the coil that are perfectly acceptable for this application but in the study presented here a coil was fitted outside the material holding cell perpendicular to the holding field and consisted of 20 turns
of nonmagnetic silver-plated-copper-clad stainless steel of $\sim$0.2 mm in diameter.  Under circumstances of limited RF power the coil can be matched and tuned.  A high RF coil Q-factor can help to minimize the needed input power but for the present application no specialized tuning was performed as power specifics were not required.  However, the change to the lineshape is sensitive to the coil Q-factor and geometry as well as orientation, so the shape change to the NMR line was studied empirically to implement in the simulation. This allows analysis in the scope of lineshape rather than the specifics of the coil and RF power.

\section{Evaluation of the Rates}
\label{eval}
\begin{figure}
\begin{center}
\includegraphics[height=50mm, angle=0]{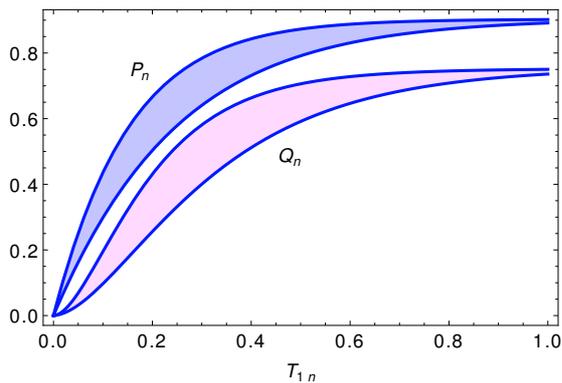}
\end{center}
\caption{The solutions to the rate equation in terms of the dynamic nuclear polarizations $P_n(\theta)$ and $Q_n(\theta)$ for spin-1 with only microwave irradiation showing the progress to the system's steady-state, using arbitrary material parameterization.  The microwave decay parameter $\beta$ is varied 10\% to make each band in the solutions.}
\label{dnp}
\end{figure}
The nonlinear coupled differential equations in Eq. \ref{rateqs2} are solved numerically to study the behavior of the family of parameterized solutions. The anharmonic Raman process is the most common nuclear-spin phonon decay leading to $T'_{1q}=\frac{5}{3}T'_{1n}$ \cite{fedders,fedders2}.  In addition, $T'_{1e} \ll T'_{1q}\sim T'_{1n}$, where $T'_{1e}$ is the parameter that sets the time scale. The free radical density parameter $c$ can be varied to study different DNP saturation
conditions for a given $T'_{1e}$ and $P_0$ allowing the study of different materials.  These characteristics can be adjusted to model the behavior and polarization states of various materials.

An initial evaluation of the rate equations for spin-1 using the low temperature limit $(r\to 0)$, $(P_0 \to -1)$ for a generic material is shown in Fig. \ref{dnp}.  The only RF-induced transitions come from the microwave source used to produce the standard DNP behavior.  Although not tuned to a real physical material, this system still exhibits the correct relationship (Eq. \ref{relation}) between vector and tensor polarization, as the nuclear Zeeman levels populate according to Maxwell-Boltzmann statistics.  The vector and tensor polarization build-up shown in
Fig. \ref{dnp}, use initial conditions $P_n(0)=Q_n(0)=0$ and $P_e(0)=-1$. The microwave decay parameter $\beta$ is varied by 10\% to generate each band. The resulting family of solutions,
previously described in \cite{fedders,jizhi}, forms the starting point for all of the extensions presented here. The units of time are in terms of the effective nuclear relaxation $T'_{1n}$, a parameter easily accessible in CW-NMR studies.

Using the rate equations to describe a particular material like ND$_3$ requires adjustment of the parameters to have the correct build-up and steady-state
saturation for a particular $T'_{1n}$ at a fixed $\beta$. In this way, the behavior of a material can be approximated for the DNP response to a particular paramagnetic center
composition.  The characteristic set of parameters specific to the material are ($T'_{1e}$, $P_0$, $c$), using the same initial conditions, such that $P_n(0)=Q_n(0)=0$ and $P_e(0)=-1$.  The set of differential equations with resulting numerical solutions are used to
fit DNP build-up from experimental data.  The parameters extracted from the fit result in $T'_{1e}=$ 35 ms,~$P_0=$-0.34,~ $c=1.8\times 10^{-4}$.  The other rates are small and have little impact on the dynamics of interest. The polarization build-up and resulting fit is shown in Fig. \ref{pcheck} with the experimental data for electron irradiated ND$_3$ that has been optimized for a nuclear scattering experiment to run at 1 K and 5 T \cite{crabb1}.

To analyze the behavior of Eq. \ref{rateqs2} at a particular $R$ in the NMR line, the steady-state intensities $I_{+}(R)$ and $I_{-}(R)$ at any frequency $R$ can be
understood by allowing the two overlapping absorption lines to increase in magnitude with respect to the Boltzmann distribution of the magnetic sublevels. The intensities at position $R=-1$
are shown in Fig. \ref{combo} (gray line) and are the results of evaluation of the theoretical lineshape of each of the absorption lines at $R$ for the modeled ND$_3$ for a vector polarization of $P_n=42\%$.
The intensities at any position $R$ in the line are defined as,
\begin{eqnarray}\nonumber
I_+(R)& =& \frac{C_E}{2}\left(P_n(R)+Q_n(R)\right)\\\nonumber
I_-(R)& =& \frac{C_E}{2}\left(P_n(R)-Q_n(R)\right).
\end{eqnarray}

To study the dynamic behavior of the system, the RF irradiation is introduced with a Heaviside step function after the DNP steady-state is reached.
The RF is turned on at $T_{1n}=2$. The progressive build-up to the steady-state polarization conditions and the changes to each intensity from the RF irradiation at position $\mathcal{R}$
are shown in Fig. \ref{intens}. The RF irradiation decay parameter, $\xi$, is more than two orders of magnitude smaller than that of $\beta$, which corresponds
to the mW scale in terms of RF power. The system approaches a new equilibrium, minimizing the negative tensor polarization $Q_n(\mathcal{R})$ at $R=-1$ resulting in a semi-saturated steady-state condition.
Equilibrium is achieved when equivalence in the loss rate from induced RF transitions and re-population rates at the position $\mathcal{R}=-1$ occurs. Using semi-saturating RF irradiation, the absorption line with the largest intensity or the greatest population difference absorbs most of the RF resulting in more induced transitions, with only marginal transitions induced in the smaller intensity absorption line.  This is advantageous for quadrupole enhancement.

\begin{figure}
\begin{center}
\includegraphics[height=50mm, angle=0]{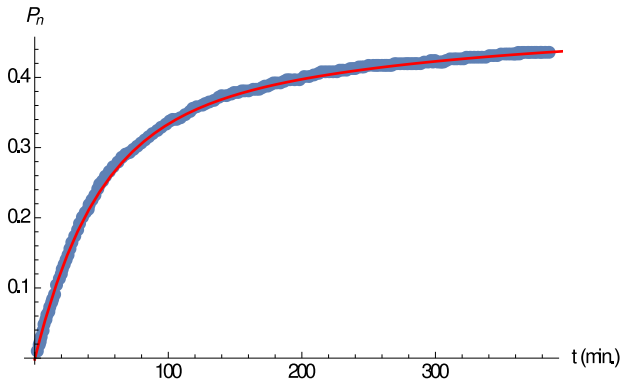}
\end{center}
\caption{The tuned solution to the rate
equations for ND$_3$ (the red line) is compared to experimental data (blue data points) for electron irradiated ND$_3$ that has been optimized for a nuclear scattering experiment to run at 1 K and 5 T.}
\label{pcheck}
\end{figure}

\begin{figure}
\begin{center}
\includegraphics[height=50mm, angle=0]{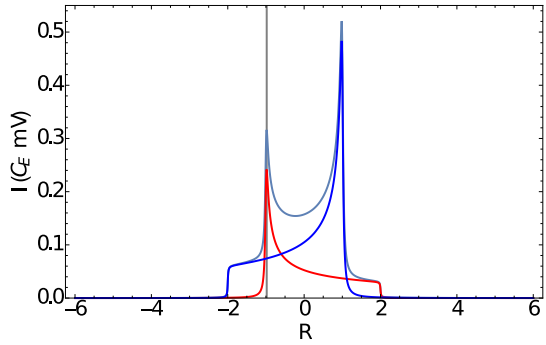}
\end{center}
\caption{The lineshape with the steady-state intensities $I_{+}(R)$ and $I_{-}(R)$. The intensities at $R=-1$ are shown as the gray vertical line
for the two overlapping absorption lines. }
\label{combo}
\end{figure}

Applying semi-saturating RF anywhere in the line reduces the absolute value of the tensor polarization quickly with respect to the effective nuclear relaxation.  The rate of minimization of $|Q_n(\mathcal{R})|$ directly results from the effective RF power. However, greater RF power results in a greater depletion of the smaller intensity which begins to work against tensor polarization enhancement. Optimal total alignment enhancement is achieved when the negative tensor polarization in the full signal has been minimized for the largest value
of the vector polarization. 
\begin{figure}
\begin{center}
\includegraphics[height=50mm, angle=0]{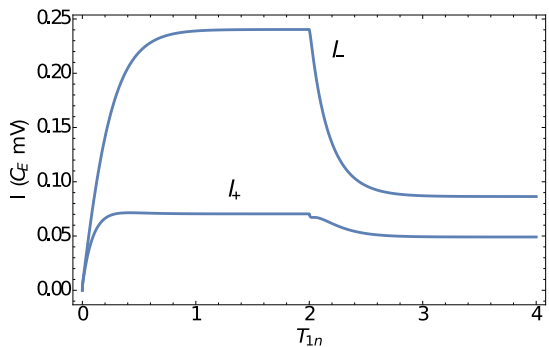}
\end{center}
\caption{The solutions to the rate equations (Eq. \ref{rateqs2}) of the resulting transition intensities after turning on RF irradiation at $T_{1n}=2$. Here the microwave decay parameter is held at 10
with the RF irradiation decay parameter $\xi$ held at 0.08.}
\label{intens}
\end{figure}

Enhancement of the tensor polarization from the minimization of $Q_n(\mathcal{R})$ has two places in the NMR line where the effect can be seen. The RF-induced transition at $\mathcal{R}$ reduces the total negative tensor polarization in the deuteron NMR signal but also increases the total positive tensor polarization by increasing the difference in the intensities at $\mathcal{-R}$, seen in both Fig. \ref{mcdis} and Fig. \ref{comp}.

\begin{figure}
\begin{center}
\includegraphics[height=55mm, angle=0]{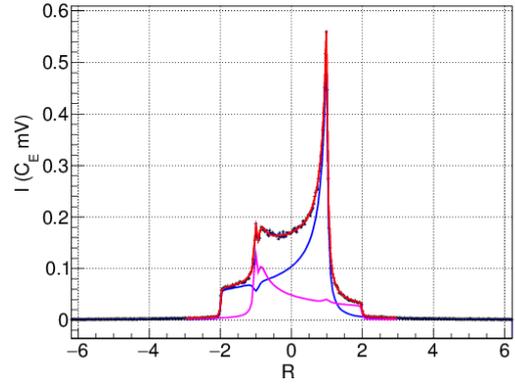}
\end{center}
\caption{The simulated lineshape (red line) based on numerical solutions to the intensities in each absorption line (pink and blue) compared to experimental data
(black points). There is a reduction in the intensities at $\mathcal{R}$ and an increase at $\mathcal{-R}$, leading to the net quadrupole enhancement in the signal.}
\label{mcdis}
\end{figure}

\section{Enhancement Optimization}
\label{opti}
To optimize the enhancement, the selective RF excitation must 
minimize the negative tensor polarization, $Q_n(R)$, for all $R$. This can be identified by looking at places in the NMR line where $I_+(R)-I_-(R)<0$. The two critical regions lie around $R\sim-1$ and $1<R<2$.
For positive vector polarization, as in Fig. \ref{fit}, the greatest integrated tensor polarization enhancement is achieved through selective excitation to reduce the size of the smaller transition area with intensity $I_-$. For negative vector polarization, the greatest enhancement comes from the reduction of the transition area with intensity $I_+$, otherwise the treatment of both cases are identical, so it is convenient to focus on positive vector polarization.
\begin{figure}
\begin{center}
\includegraphics[height=55mm, angle=0]{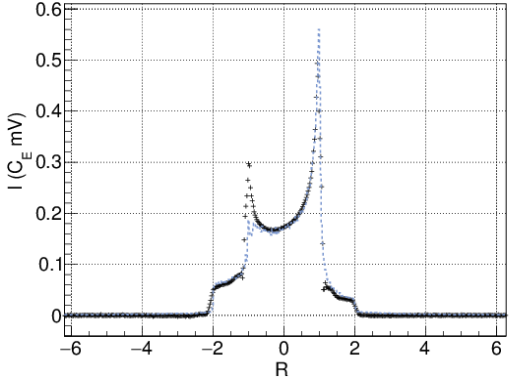}
\end{center}
\caption{Change in total intensity showing before RF irradiation and after the semi-saturated steady-state is reached. The black points are before and the
dashed line is after. Both lines are from experimental NMR data.}
\label{comp}
\end{figure}

Minimization of the smaller transition area ($I_-$ in this case) around $1<R<2$ is managed using continuously modulated RF with sufficient power to lead to
full saturation, equalizing the three sublevels in the irradiated frequency domain. The region around $R=\pm1$ does not benefit as much by full saturation due
to the significant overlap with the other transition ($I_+$ in this case).

The rate equations are used to model the simultaneous change of transition intensities at the irradiated NMR line positions. The results of the dynamic solutions for
the overlapping absorption lines define the change in amplitude of the Voigt function (from Section \ref{rates}) at $\mathcal{R}$. The rate equations are 
parameterized to represent the polarization state of an NMR signal unique to position $\mathcal{R}$ in the line under the RF irradiation. The initial difference in the absorption lines is set using the normalization constants $g_i$.  This initial condition is easy to calculate being the system starts in Boltzmann equilibrium. The simulation incorporates the intensity changes from the numerical solution with the Voigt shape from data resulting in a generated CW-NMR signal.  Figure \ref{mcdis} shows the
simulated lineshape (red line) based on the numerical solutions of the intensities in each absorption line (pink and blue) in comparison to ND$_3$ experimental data at 1 K and 5 T, seen as the black points.  In this example the RF is applied to a single position at the smaller peak.
This example serves as a qualitative check between simulation and experimental data in matching the sum of intensities.  Figure \ref{comp} shows a comparison of the experimental NMR lineshape before and after the RF irradiation. 
The distribution of negative tensor polarization changes about the $R=0$ axis for negative vector polarization. Analytical minimization is easily seen for the positive case by studying the tensor polarization as a function of frequency. An example of the tensor polarization as a function of line position $Q_n(R)$ is shown in Fig. \ref{quad} for a vector polarization of 42\%.
\begin{figure}
\begin{center}
\includegraphics[height=50mm, angle=0]{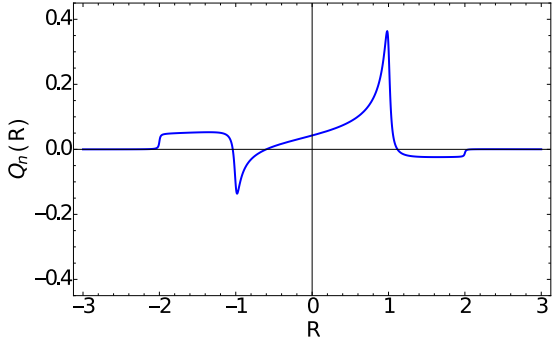}
\end{center}
\caption{The tensor polarization as a function of line position $Q_n(R)$ for a vector polarization of 42\%. Optimization of the tensor polarization is obtained
by negating the values that drop below zero using RF induced transition for that selected frequency range.}
\label{quad}
\end{figure}
The necessary RF irradiation-induced decay constant, or alternatively the required RF coil power, can be calculated for the minimization of $Q_n(R)<0$ for overlapping transition lines by finding the extrema for the expression for partial saturation efficiency,
\begin{equation}
\Phi(\xi)= \lim_{t \to \infty} P_n\int^{\infty}_{-\infty} Q^+_n(R)+Q^-_n(\mathcal{R})dR.
\end{equation}
The efficiency $\Phi(\xi)$ applies to the steady-state conditions for a value of RF decay constant $\xi$ with DNP microwaves active.  The value of $\xi$ can be varied at each position $\mathcal{R}$ where the tensor polarization is negative ($Q^-_n$)
to achieve the maximum alignment for a given vector polarization $P_n$.  The efficiency is greatest for a integrated alignment where the positive tensor polarization ($Q^+_n$) is large and the negative tensor polarization is mitigated with only minimal reduction of $P_n$ in the process. The critical $\xi$ value for optimization can be calculated using,
\begin{equation}
\lim_{t \to \infty}\frac{d\Phi(\xi_c)}{d\xi}=0.
\end{equation}
The steady-state critical value $\xi_c$ depends on the microwave power parameter $\beta$, the composition of paramagnetic centers $c$, the relaxation rates of the system, the
temperature and holding field strength, and the position in the line.

There is a steady-state critical value $\xi_c$ at every position in the NMR line for $Q_n^-$. For optimal $Q_n$ enhancement, the modulated RF changes in central frequency and $B_\nu$ to meet the required $\xi_c$ for each position. Given the initial RF response parameterization the theoretical lineshape can be used to achieve the optimal $Q_n$ enhancement, for any initial vector polarization, without a direct measurement of the effective RF power.

The theoretical optimization for the CW-NMR lineshape is shown in Fig. \ref{fin}. The simulation is used to generate a CW-NMR signal for the deuteron target sample with initial nuclear
polarizations of 12\%, 42\%, and 78\%. The simulation then responds to RF irradiation at the line position $\mathcal{R}$ with an RF decay constant $\xi_c$, changing as required to minimize
$Q_n^-$.  Full enhancement occurs when the irradiated intensities reach a steady-state for both the saturated region $1<R<2$ and the partially saturated region $R=-1$. The final alignment-optimized steady-state lineshape appears in dark blue. The simulation study results in the maximum tensor polarizations of $1.3\to5.4\%$, $13.6\to23.8\%$, and $52.2\to58.5\%$, respectively.  These values represent the theoretical maximums for various vector polarizations for a static polycrystalline target with all discussed contributions. Additional enhancement is possible in slowly rotating targets, which later works will address.

In electron-scattering experiments where the target uses a 1 K evaporation refrigerator and a 5 T holding field, ND$_3$ is realistically expected to polarize to around 42\% at the dose where the materials paramagnetic centers are optimized and with the heat load of the electron beam. Under optimal condition, with the beam off, ND$_3$ can polarize significantly more ($\ge$50\%), however it takes much longer than most experiments' allotted polarization build-up time (greater than 10 hours using 1 W microwave power on the target).
\begin{figure}
\begin{center}
\includegraphics[height=50mm, angle=0]{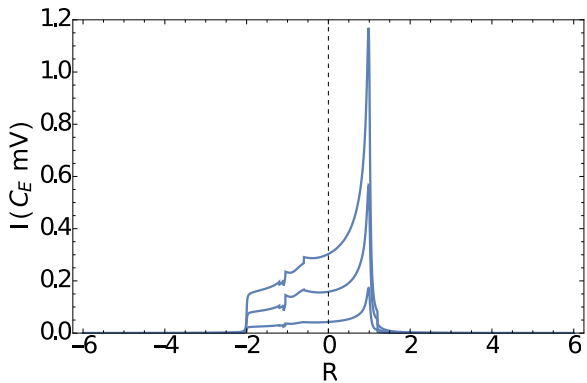}
\end{center}
\caption{The theoretical CW-NMR lineshape for optimal quadrupole enhancement at the semi-saturation steady-state condition for each polarization shown in Fig. \ref{lines}. The plots are generated with a simulation of the deuteron signal when RF irradiation is applied to the frequencies with negative tensor polarization with optimal RF power ($\xi_c$) for the overlapping region. }
\label{fin}
\end{figure}

\section{Experimental Verification}
\label{veri}
Using the simulation and the theoretical lineshape for the semi-saturated steady-state, it is possible to predict the enhanced deuteron target alignment achievable during nuclear and particle physics experiments.  However, the comparisons here between CW-NMR data and the predictions are not independently conclusive. The comparisons only confirm that
the sum of the two absorption lines remains consistent with the calculations presented here, while leaving room for error regarding their difference, particularly at the position of RF irradiation.  The difference of the lines is needed for an unambiguous verification of the tensor polarization.
 The simulation can continue to be improved using empirical information from solid-state experiments.  More precise parameterization of the diffusion constant and the RF-induced transition rate as a function of $\theta$ can be studied using specialized solid-state NMR experiments. One possibility is to use RF selective saturation at a particular $\theta$ and then slightly rotate the sample while tilting at the magic angle to separate and measure the intensity change in each absorption line. This type
of simulation can be used to study our understanding of many interdependent rate dynamics and the resulting polarization in the target. For these results to be useful for CW-NMR measurement, it is necessary to develop a full-fitting technique which uses the theoretical lineshape discussed.

The most unambiguous verification of the tensor polarization measurement is through nuclear scattering experiments. In particular, in JLab experiment E12-14-002, the kinematic requirements of the tensor asymmetry in the quasi-elastic region allows for the simultaneous measurement of elastic $T_{20}$ at multiple $Q^2$ points ranging from $0.2<Q^2<1.8$ GeV$^2$. In the region
$Q^2\sim0.2$ GeV$^2$, the $T_{20}$ measurements from world data \cite{Holt,Zhang,Luzzi,Schulze} can be used to determine the tensor polarization of the target to within less than 6\%, using the proposed precision for this calibration point. This single measurement can verify and set the uncertainty of the tensor enhanced theoretical lineshape presented here. Once verified, the discussed measurement technique is required for all other kinematic settings for E12-14-002 and E13-12-011, as well as possible future solid tensor polarized target experiments such as tensor polarized Drell-Yan \cite{kum1,kum2,kum3} and tensor polarized DVCS \cite{kirch,air}.

Additional scattering experiments can be used to reduce uncertainty in the lineshape tensor polarization measurements. Additional tensor polarized observables can be exploited through changes in orientation of the quantization axis. The target tensor asymmetry
decomposition into analyzing powers is expressed as,
\begin{eqnarray}
A^T_d&=&\frac{3\cos^2\theta^*-1}{2}T_{20}-\sqrt{\frac{3}{2}}\sin 2\theta^*\cos\phi^*T_{21}\\\nonumber
&+&\sqrt{\frac{3}{2}}\sin^2\theta^*\cos 2\phi^*T_{22},
\end{eqnarray}
where the density matrix, Eq. \ref{one}, maintains the same symmetry but the polarization direction is now described by the polar and azimuthal angles $\theta^*$ and $\phi^*$ in a frame where
the $z$-axis lies along the direction of the virtual photon and the $y$-axis is defined by the vector product of the incoming
and outgoing electron momenta. With the target polarization oriented in this fashion, it is possible to selectively study each analyzing power where world data can
be used to minimize uncertainties. These types of studies need not be limited to electron scattering. Photodisintegration of the deuteron
also provides a significant amount of world data \cite{Holt,Ranchek,Mishnev} which could potentially be used if a solid target was alignment enhanced following the optimization suggested. 

\section{Conclusion}
\label{con}
To achieve the highest FOM for the spin-1 alignment scattering experiments, it is necessary to maximize the tensor polarization of the solid-state targets throughout the beam-target interaction time. Optimization of the spin-1 alignment in the target ensemble can be achieved by applying RF irradiation at select frequencies. A simple theoretical lineshape has been introduced that is based on empirical information and the numerical solutions to the solid-effect rate equations. The solutions allow
determination of the steady-state intensities at the position in the line under semi-saturation from RF irradiation. A simulation parameterized by fitting to the polarization build-up rate is developed and used to study the dynamics of the two absorption lines in the CW-NMR.  Optimization is achieved using a critical RF-driven decay constant that partially saturates
the overlapping regions of the Pake doublet corresponding to $Q_n(R)<0$. Enhancement results from reducing negative contributions to the tensor polarization while simultaneously increasing positive contributions.  RF manipulation of the population of the $m=0$ magnetic substate is observed in the NMR signal as a decrease in the intensities of the transitions at the RF line position and an increase in opposing transitions at the same $\theta$. Qualitative comparison of the simulated ND$_3$, CW-NMR signal to real experimental data under similar conditions shows very good matching in overall intensity. A tensor polarization of $\sim$24\% is predicted given an averaged vector polarization of 42\% in a scattering experiment.

For any given vector polarization, there exists a lineshape that optimizes tensor polarization enhancement, accessible using RF irradiation at select positions in the deuteron NMR line.  Given the initial RF response parameterization, the theoretical lineshape discussed can be used to achieve optimal $Q_n$ enhancement, for any initial vector polarization, without a direct measurement of the effective RF power.
New constraints based on the optimized RF steady-state intensities can be used to fit and measure the manipulated CW-NMR line over the course of the scattering experiment to monitor the degree of tensor polarization.  Later work will present this.

It is possible to take these enhancement and measurement techniques and apply them to slow target rotation for additional enhancement.  However, for a rotating target the level of enhancement achievable is very much material dependent.  Work on deuterated butanol and ND$_3$ is presently underway.  It has also been suggested that single crystals of ND$_3$ could be used so that a greater percentage of NMR area pertaining only to negative tensor polarization can be saturated \cite{bruss}.  The usage of single crystals may be beneficial specifically with photon beams where beam heating and radiation damage from the charged beam is not as problematic.

The optimization approach presented here provides the required formalism and details for the simulation needed for predictions of tensor polarization enhancement of the spin-1 target. 
Nuclear scattering experiments can verify these predictions. Complete verification of this characterization of deuterated ammonia allows for the development of future tensor-polarized experiments with the most essential 
solid spin-1 target compatible with charged particle beams.

\section*{Acknowledgement}

The author thanks D. Crabb and C. Keith for a thorough read.
Numerical solutions are produced with Wolfram Research, Inc., Mathematica, Version 10.0. All NMR data presented here was taken at the University of Virginia Solid Polarized Target Lab. This work was supported by DOE contract DE-FG02-96ER40950.

\end{document}